\documentclass[10pt, a4paper]{article}

\usepackage[T1]{fontenc}
\usepackage[utf8]{inputenc}
\usepackage{lmodern}          
\usepackage[scaled]{helvet}   

\usepackage{geometry}
\usepackage{setspace}
\usepackage{titlesec}
\titleformat{\section}
  {\sffamily\fontsize{12}{14}\selectfont\bfseries}
  {}{0em}{}
\titlespacing{\section}{0pt}{12pt}{6pt}

\titleformat{\subsection}
  {\sffamily\fontsize{10}{12}\selectfont\bfseries}
  {}{0em}{}
\titlespacing{\subsection}{0pt}{10pt}{4pt}

\titleformat{\subsubsection}
  {\sffamily\fontsize{10}{12}\selectfont\itshape}
  {}{0em}{}
\titlespacing{\subsubsection}{0pt}{8pt}{0pt}

\usepackage{orcidlink} 
\usepackage{graphicx}
\usepackage{booktabs}
\usepackage{xcolor}
\usepackage{amsmath}
\usepackage{hyperref}
\usepackage{enumitem}           
\setlist[itemize]{noitemsep, topsep=4pt}

\usepackage[
  style=apa,
  sortcites=true,
  sorting=nyt,
  backend=biber
]{biblatex}
\DeclareLanguageMapping{english}{english-apa}
\let\cite\parencite
\makeatletter
\renewcommand{\maketitle}{%
  {\sffamily\fontsize{14}{16}\selectfont\bfseries\noindent
    The GenAI Skill Bypass: Mapping Divergent Pathways of University
    Students and Staff AI Literacy
  \par}
  \vspace{10pt}
  {\noindent\normalsize
    Eduardo Oliveira\textsuperscript{1}*\orcidlink{0000-0001-5063-8860},
    Narelle English\textsuperscript{1}\orcidlink{0000-0002-4726-560X},
    Tracii Ryan\textsuperscript{1}\orcidlink{0000-0002-0512-5713},
    Kamila Misiejuk\textsuperscript{2}\orcidlink{0000-0003-0761-8703},
    Cory dal Ponte\textsuperscript{1}\orcidlink{0009-0007-4444-813X},
    Sonsoles López-Pernas\textsuperscript{3}\orcidlink{0000-0002-9621-1392}, and
    Mohammed Saqr\textsuperscript{3}\orcidlink{0000-0001-5881-3109}
  \par}
  \vspace{1pt}
  {\noindent\small
    \textsuperscript{1}University of Melbourne, Australia \\
    \textsuperscript{2}FernUniversit\"{a}t, Germany \\
    \textsuperscript{3}University of Eastern Finland, Finland
  \par}
\vspace{1pt}
  {\noindent\small
    \textit{*Corresponding author:} Eduardo Oliveira
    (\href{mailto:eduardo.oliveira@unimelb.edu.au}{eduardo.oliveira@unimelb.edu.au})
  \par}
  \vspace{10pt}
\begin{list}{}{%
  \setlength{\leftmargin}{1.0cm}
  \setlength{\rightmargin}{1.0cm}
  \setlength{\topsep}{0pt}
}
\item\small\itshape
  This is a Preprint. This manuscript is currently under review and has
  not yet been peer-reviewed or accepted for publication.
\end{list}
}
\makeatother

\begin{document}

\maketitle

\begin{list}{}{%
  \setlength{\leftmargin}{1.0cm}
  \setlength{\rightmargin}{1.0cm}
  \setlength{\topsep}{10pt}
}
\item
Higher education institutions are increasingly expected to ensure that both students and staff develop Generative AI (GenAI) literacies. In response, they are introducing professional development programs and embedding GenAI skills within student curricula. However, current educational frameworks typically assume a linear progression of GenAI literacy, implying that foundational technical understanding must precede creative application. This paper challenges such an assumption through a psychometric analysis of a taxonomy-based self-assessment instrument ($n$ = 158). We applied Rasch measurement theory and Guttman ordering to map the latent perceived order of difficulty of GenAI skills across students, academics, and professional staff. Results reveal a fundamental divergence in perceived competence profiles: while academics follow a more traditional linear path, students exhibit an ``inverted'' profile, frequently mastering high-level creation tasks before acquiring foundational conceptual understanding. Furthermore, the correlation of skill difficulty between students and academics was weak ($r$ = 0.188). We argue that this ``skill bypass'' creates a fragile sense of fluency, where high self-efficacy in prompting masks low literacy in AI mechanics. These findings challenge the ``one-size-fits-all'' curricula and provide the empirical basis for diagnostic-driven, modular interventions that foster genuine human-AI synergy.
\end{list}

\vspace{10pt}
\textit{Implications for practice or policy}

\begin{itemize}
  \item Curriculum designers should not assume students learn GenAI skills foundation-first, since this study reveals that many master AI-assisted creation before conceptual, ethical, and evaluative understanding.
  \item Educators could consider treating early student fluency as fragile, embedding evaluative and ethical understanding into AI workflows to counter over-reliance.
  \item Institutional leaders could implement training based on modular pathways differentiated for students, academics, and professional staff, rather than deploying one-size-fits-all approaches.
  \item Researchers should continue modelling AI literacy as a role-dependent, non-linear construct rather than a single uniform developmental ladder.
\end{itemize}

\vspace{10pt}

\section{Introduction}
The integration of Generative AI (GenAI) into higher education has shifted from a phase of rapid experimentation to a pressing demand for sustainable, evidence-based and responsible implementation. As GenAI tools become ubiquitous in professional and academic workflows, the primary challenge for institutions is no longer merely providing staff and students with access, but bridging the critical ``competency gap'' between basic exposure and true fluency. For example, Australia's higher education regulator, the Tertiary Education Quality and Standards Agency (TEQSA), recently encouraged institutions to take a whole-of-institution approach to developing AI literacy, to ensure ethical and appropriate usage of these tools and enable students to meet the learning outcomes of their program. However, as recent frameworks emphasise, navigating this landscape requires a comprehensive, multidimensional literacy that integrates theoretical knowledge, practical application, and ethical reflection \cite{bozkurt2024generative}. Such fluency is a prerequisite for learners (students, academics, and professional staff alike) to act as capable agents in a human-AI loop, rather than passive consumers of output.

Recent literature, particularly in higher education, has successfully defined the necessary competencies for the use of GenAI, ranging from prompt engineering to ethical reasoning. Some scholars have also proposed theoretical roadmaps for how these skills should be sequentially acquired \cite{annapureddy2025generative}, and developed robust performance-based metrics \cite{jin2025glat}. However, the field lacks empirical evidence regarding the developmental trajectories in which learners acquire these skills. Specifically, institutions who aim to develop the AI literacy of their students and staff face a practical implementation challenge: while the competencies required for GenAI literacy are increasingly well-defined, the sequence in which different university community members (students, academics, and professional staff) actually acquire these capabilities remains empirically uncharted. A key theme emerging from recent discourse within the GenAI research is the necessity to move beyond theoretical frameworks toward empirical validation in real-world settings \cite{kasneci2023chatgpt}.

This gap is compounded by a mismatch between assumed and actual learning sequences. Existing GenAI literacy models often imply a linear progression of learning, assuming that foundational technical knowledge must precede creative application or ethical evaluation \cite{krathwohl2002revision}. For instance, recent instructional paradigms strongly advocate for scaffolded AI integration aligned with Bloom's Taxonomy, arguing that learners must independently demonstrate proficiency at each cognitive level before being permitted to leverage generative systems to extend or synthesise their work \cite{hutson2025scaffolded}. However, the accessible nature of GenAI interfaces invites an inversion of this pedagogical sequence. Recent performance-based studies suggest that learners often possess scattered profiles of competence; for example, students may be proficient in utilising chatbots for content generation (creation) but lack the critical literacy to evaluate hallucinations or copyright implications (understanding) \cite{jin2025glat,lyu2024evaluating}. This disconnect poses a risk to the formation of effective Human-AI teams, as users may over-trust their own capability based on the ease of producing outputs.

Understanding these real developmental trajectories has direct consequences for how learners engage with upskilling. Sustained engagement in learning is heavily influenced by a sense of autonomy and competence \cite{ryan2000self}; when institutional curriculum design (e.g., through professional development modules) imposes rigid, linear pathways that do not match learners' actual skill levels, motivation often wanes. To foster agency, learners must be able to self-diagnose their own capability gaps and navigate modular learning pathways accordingly \cite{blaschke2012heutagogy}. Bridging the gap from framework to practice requires transitioning from abstract competency lists to actionable, diagnostic instruments. To be effective, such instruments must ground AI competencies within established pedagogical structures, such as Bloom's taxonomy, to support self-determined learning \cite{krathwohl2002revision,blaschke2012heutagogy}.

While recent theoretical models have attempted to map AI competencies onto established taxonomic frameworks, most notably the GenAI Literacy and Fluency framework \cite{dal2025scaffolding}, these designs have largely remained conceptual. The critical next step for the field is to subject such theoretical models to rigorous empirical scrutiny. This paper addresses this imperative through two connected aims. First, we empirically examine whether a taxonomy-based self-assessment instrument ($n=158$) measures a coherent construct of GenAI literacy across a diverse university population --- moving beyond theoretical design to investigate its psychometric properties. Second, and consequentially, we investigate how the specific developmental trajectories of GenAI skills differs across distinct university cohorts (students, academics, and professional staff), on the premise that effective GenAI implementation requires understanding not only \textit{what} competencies learners need, but \textit{when} and in what sequence they actually develop them. As such, we applied Rasch measurement theory \cite{masters1982rasch} and Guttman ordering analysis \cite{guttman1950problem,guttman1954principal} to investigate the ``order of emergence'' of GenAI skills. Unlike traditional validation studies that seek a single fit for all users, our results demonstrate that GenAI literacy is not a uniform developmental path; rather, students and staff follow distinct trajectories based on their pragmatic needs and contexts. These findings reframe the challenge of GenAI literacy development from a content problem --- \textit{what} to teach --- to a sequencing problem --- \textit{where} each learner actually stands. By mapping role-specific orders of emergence, this study provides both the empirical rationale and a diagnostic pathway for institutions to replace uniform curricula with targeted, modular interventions that build from learners' existing capability profiles rather than assumed starting points.
\section{Background and Literature Review}

The rapid proliferation of GenAI requires a paradigm shift in how digital literacies are conceptualised, taught, and measured in higher education. Unlike passive interactions with traditional machine learning, GenAI requires a distinct ``fluency'' characterised by iterative interaction, prompt engineering, and critical evaluation of probabilistic outputs. Recent multidimensional models argue that this fluency must balance foundational knowledge (`Know What'), practical application (`Know How'), and critical, ethical reflection (`Know Why') \cite{bozkurt2024generative}. A systematic review \cite{almatrafi2024systematic} synthesised this landscape into six core constructs: Recognise, Know \& Understand, Use \& Apply, Evaluate, Create, and Navigate Ethically. Although this taxonomy provides a valuable vocabulary, current educational interventions significantly imbalance these domains by prioritising lower-order ``knowing'' and ``understanding'' while neglecting higher-order ethical navigation \cite{almatrafi2024systematic}. Essentially, GenAI literacy cannot be decoupled from domain expertise. Even though experts integrate AI skills with deep subject knowledge to assess accuracy of outputs, novices often lack the domain expertise required to evaluate hallucinations, creating a dependency risk. This learner behaviour mirrors what Mollick \cite{mollick2025jagged} describes as the ``Jagged Frontier'' of AI capability, where peaks of sophisticated performance sit alongside surprising valleys of basic incompetence. Recent studies in medical education confirm this architectural quirk, finding that LLM errors often cluster at the lower rungs of Bloom's ladder (remembering and applying) despite high performance in synthesis \cite{herrmann2024assessing,galla2025generative}.

A central tension in AI research lies in how best to measure these complex literacies. The field is divided between self-reported measures, capturing learner confidence, and performance-based measures, capturing demonstrable skill. Recent scholarship has critiqued reliance on self-reports due to the ``illusion of explanatory depth'' \cite{laupichler2023development,ng2021conceptualizing}. In response, \textcite{jin2025glat} developed the GenAI Literacy Assessment Test (GLAT), arguing that performance-based tasks are a more reliable indicator of technical capability. However, distinct pedagogical functions exist between these approaches. While performance testing is the gold standard for summative verification, it differs fundamentally from diagnostic self-assessment \cite{boud2000sustainable}. To act as an effective teammate to an AI agent, a human must possess agency. \textcite{bearman2024developing} posit that the critical human capability is ``evaluative judgement''--the ability to discern the quality of work of self and others. Because GenAI outputs are prone to subtle errors, learners must develop the metacognitive habit of auditing the technology \cite{walton2025university}. From this perspective, assessment must do more than verify competence; it must cultivate agency by allowing learners to self-diagnose gaps \cite{blaschke2012heutagogy}.

While the ``what'' (competencies) and ``how'' (assessment) of GenAI literacy are defined, the ``when''--the developmental sequence of skill acquisition--remains a critical blind spot. Educational frameworks grounded in Bloom's taxonomy are often interpreted as prescribing a linear sequence in curriculum implementation, despite \textcite{krathwohl2002revision} acknowledgment that higher-order skills may be engaged in varying orders depending on task context. Indeed, contemporary pedagogical models advocate for strict scaffolded integration, arguing that learners must independently demonstrate foundational comprehension before being permitted to leverage generative systems for higher-order synthesis and creation \cite{hutson2025scaffolded}. However, the accessible nature of GenAI interfaces invites an inversion of this sequence; users often engage in high-level \textit{creation} and \textit{application} without possessing the foundational \textit{understanding} of the underlying technology. \textcite{galla2025generative} term this phenomenon ``Bloom's Inversion,'' arguing that GenAI architecture is fundamentally optimised for higher-order pattern generation (\textit{creation}) while frequently failing at lower-order factual recall (\textit{remembering}). This disconnect suggests that GenAI literacy may not develop in the uniform, linear trajectory assumed by curriculum designers. Instead, skill acquisition is likely driven by pragmatic necessities: students may refine \textit{creative} prompting strategies to complete assessments long before they engage with \textit{ethical} data privacy guidelines \cite{lyu2024evaluating}, while professional staff may prioritise regulatory compliance before technical creation. Consequently, theoretical models are beginning to propose a ``Reversed Bloom's'' pedagogy, where the cognitive burden shifts from initial generation to the critical evaluation and refinement of AI-produced artifacts \cite{pesovski2024ai}.

A clarification is needed on how we operationalise ``creation.'' In the revised Bloom's taxonomy, Create denotes synthesising elements into a novel whole, with the learner as generative agent \cite{krathwohl2002revision}. When a student prompts a GenAI tool, the model performs the generative act while the student specifies the task, evaluates outputs, and refines iteratively---so we do not claim this is cognitively equivalent to unaided Create. As suggested by \textcite{galla2025generative}, we instead treat Create-level engagement within GenAI workflows as the capacity to formulate effective prompts, critically assess probabilistic outputs, and integrate generated artefacts into purposeful products---a higher-order but distinct skill we term \textit{functional creation}. The ``skill bypass'' lies precisely here: students develop functional-creation fluency while bypassing the foundational understanding needed to audit, correct, and contextualise the AI's output.

\begin{figure}[!t]
    \centering
    \includegraphics[width=\textwidth]{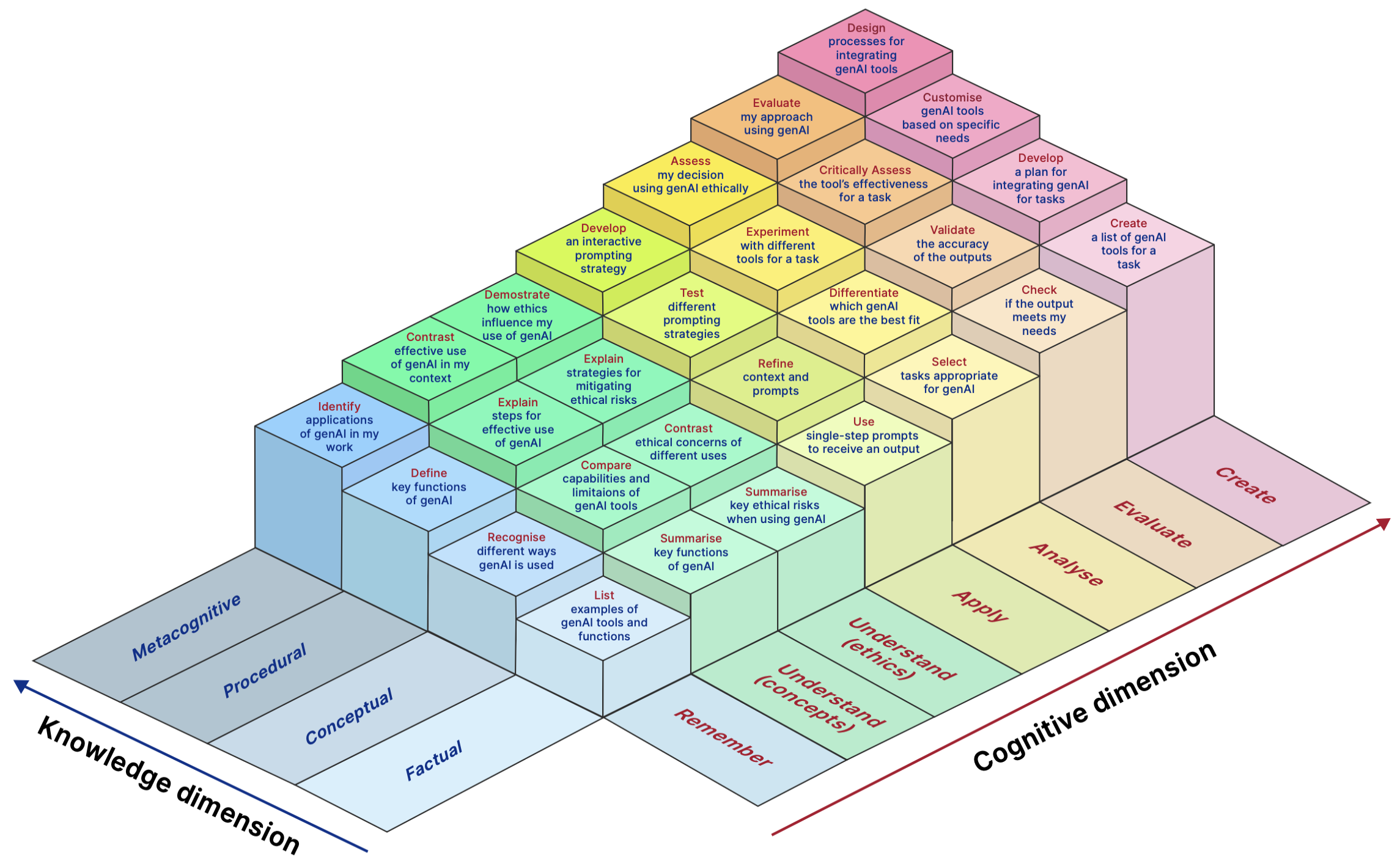} 
    \caption{Three-dimensional visualisation of a 28-item GenAI literacy and fluency self-assessment instrument (proposed by \textcite{dal2025scaffolding})}
    \label{fig:matrix}
\end{figure}

Several GenAI literacy frameworks have been proposed in recent years. For example, \textcite{annapureddy2025generative} recently proposed a 12-competency model structured as a logical developmental continuum, moving from basic AI awareness to advanced prompt engineering and ethical application. \textcite{choi2025genai} synthesised seven existing frameworks to identify key GenAI competencies for students, organising them across four levels of Bloom's taxonomy (\textit{understand}, \textit{apply}, \textit{analyse \& evaluate}, \textit{create}). While these syntheses provide a high-level competency overview, they offer limited operational detail. \textcite{dal2025scaffolding} proposed a more fine-grained mapping of GenAI skills to Bloom's taxonomy, the GenAI Literacy and Fluency framework, that includes 28 behavioural indicators across cognitive and knowledge domains. Represented as a two-dimensional matrix, the framework supports the identification of developmental learning objectives and learner needs, which offers a clear and actionable alternative to more abstract competency frameworks (see Figure~\ref{fig:matrix}). Due to its level of detail and pedagogical rigour, this framework was adopted as the theoretical foundation for this study.

Psychometric theory posits that complex competencies form a latent developmental hierarchy, or an ``order of emergence'' \cite{masters1982rasch}. Yet, we lack empirical evidence regarding what this order looks like for GenAI. If we assume a single theoretical path, we risk designing interventions that fail to meet users at their point of need. By applying probabilistic measurement models, this study aims to empirically validate these developmental pathways, determining whether the theoretical progression of GenAI skills holds true in practice or varies significantly across academic cohorts.

Empirically investigating this order of emergence requires psychometric methods capable of modelling both population-level trait structure and subgroup-level response patterns; the analytic framework for doing so is detailed in Section~\ref{methodology}.
\section{Current Study}

This study moves beyond the theoretical design of the GenAI Literacy and Fluency framework \cite{dal2025scaffolding} to test, with real-world data, its central premise: that learners can meaningfully self-diagnose their capabilities across a developmental spectrum. Rather than a validation exercise, we use the framework as a diagnostic probe to map the latent structure of GenAI literacy and to test whether the linear, ``one-size-fits-all'' progression assumed by traditional taxonomies holds across distinct university cohorts (students, academics, and professional staff). We address two research questions:
\begin{itemize}
    \item RQ1: To what extent do the instrument's items function as indicators of a shared underlying construct across a diverse university population, and does that population exhibit homogeneous or heterogeneous response patterns?
    \item RQ2: What are the developmental trajectories (``order of emergence'') of GenAI skills across cohorts (e.g., students versus academics), and what do these pathways reveal about their distinct learning needs?
\end{itemize}

\section{Methodology}
\label{methodology}
\subsection{Participants}

The sample for this study, which comprised \textit{N} = 158 respondents to an online survey, included two primary sub-groups which are central to this study's comparative analysis: 98 university students (62\%) and 60 staff members (38\%).  The staff group comprised both professional ($n = 26$) and academic ($n = 34$) staff. The participant group was diverse in age and gender (Table \ref{tab:demographics}). The largest age group was 18--24 years (41.1\%), followed by 25--34 years (25.9\%). The majority of respondents identified as female (58.2\%), with 39.9\% identifying as male. The student cohort was primarily composed of individuals enrolled in Master's degrees (60.2\%) and Bachelor's degrees (26.5\%). The staff cohort demonstrated a high level of educational attainment, with a majority holding Master's (41.7\%) or Doctoral (33.3\%) degrees. 

\begin{table}[!t]
\centering
\caption{Participant demographics by subgroup ($N = 158$)}
\label{tab:demographics}
\small
\begin{tabular}{p{4.2cm} r r r r}
    \toprule
    & \textbf{Students} & \textbf{Academics} & \textbf{Prof.\ Staff} & \textbf{Total} \\
    & \textit{n = 98} & \textit{n = 34} & \textit{n = 26} & \textit{n = 158} \\
    \midrule
    \multicolumn{5}{l}{\textit{Age group}} \\
    \quad 18--24          & 62 (63.3\%) &  2 (5.9\%)  &  1 (3.8\%)  & 65 (41.1\%) \\
    \quad 25--34          & 28 (28.6\%) &  6 (17.6\%) &  7 (26.9\%) & 41 (25.9\%) \\
    \quad 35--44          &  4 (4.1\%)  & 11 (32.4\%) &  8 (30.8\%) & 23 (14.6\%) \\
    \quad 45--54          &  3 (3.1\%)  &  8 (23.5\%) &  6 (23.1\%) & 17 (10.8\%) \\
    \quad 55--64          &  0 (0.0\%)  &  6 (17.6\%) &  4 (15.4\%) & 10 (6.3\%)  \\
    \quad 65 or older     &  1 (1.0\%)  &  1 (2.9\%)  &  0 (0.0\%)  &  2 (1.3\%)  \\
    \addlinespace
    \multicolumn{5}{l}{\textit{Gender}} \\
    \quad Female          & 52 (53.1\%) & 20 (58.8\%) & 20 (76.9\%) & 92 (58.2\%) \\
    \quad Male            & 45 (45.9\%) & 12 (35.3\%) &  6 (23.1\%) & 63 (39.9\%) \\
    \quad Non-binary      &  0 (0.0\%)  &  1 (2.9\%)  &  0 (0.0\%)  &  1 (0.6\%)  \\
    \quad Prefer not to say &  1 (1.0\%) &  1 (2.9\%) &  0 (0.0\%)  &  2 (1.3\%)  \\
    \addlinespace
    \multicolumn{5}{l}{\textit{Enrolment / highest qualification\textsuperscript{a}}} \\
    \quad Bachelor's degree       & 26 (26.5\%) &  1 (2.9\%)  &  8 (30.8\%) & 35 (22.2\%) \\
    \quad Master's (coursework)   & 59 (60.2\%) & 11 (32.4\%) &  9 (34.6\%) & 79 (50.0\%) \\
    \quad Master's (research)     &  2 (2.0\%)  &  3 (8.8\%)  &  2 (7.7\%)  &  7 (4.4\%)  \\
    \quad Grad.\ cert.\ / diploma &  1 (1.0\%)  &  1 (2.9\%)  &  5 (19.2\%) &  7 (4.4\%)  \\
    \quad Doctoral (coursework)   &  6 (6.1\%)  &  2 (5.9\%)  &  0 (0.0\%)  &  8 (5.1\%)  \\
    \quad Doctoral (research)     &  4 (4.1\%)  & 16 (47.1\%) &  2 (7.7\%)  & 22 (13.9\%) \\
    \addlinespace
    \multicolumn{5}{l}{\textit{GenAI usage frequency}} \\
    \quad Daily           & 26 (26.5\%) &  9 (26.5\%) &  5 (19.2\%) & 40 (25.3\%) \\
    \quad Most days/week  & 46 (46.9\%) & 15 (44.1\%) & 11 (42.3\%) & 72 (45.6\%) \\
    \quad Few times/month & 21 (21.4\%) &  6 (17.6\%) &  8 (30.8\%) & 35 (22.2\%) \\
    \quad Few times/year  &  5 (5.1\%)  &  4 (11.8\%) &  2 (7.7\%)  & 11 (7.0\%)  \\
    \bottomrule
    \multicolumn{5}{p{13.5cm}}{\footnotesize
        \textit{Note.} Grad.\ cert.\ = Graduate Certificate / Professional Certificate / Graduate Diploma. \textsuperscript{a}~For students, qualification refers to current enrollment level; for academic and professional staff, it refers to highest attained qualification.}
\end{tabular}
\end{table}

\subsection{GenAI Self-Assessment Instrument}

\begin{figure}[ht]
    \centering
    \includegraphics[width=0.9\textwidth]{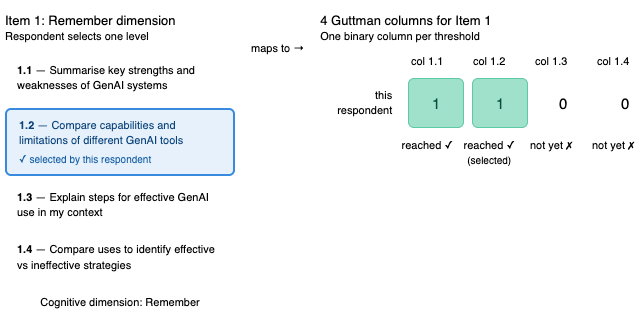} 
    \caption{Illustration of how a single polytomous item maps  to four binary threshold columns in the Guttman chart, using Item~1 (Remember cognitive dimension) as an example. A respondent selecting level~1.2 receives a score of 1 in columns for all thresholds at or below the selected level (1.1 and 1.2) and 0 in all columns above it (1.3 and 1.4). This cumulative coding rule applies across all 7 items, yielding 28 binary columns per respondent (4 thresholds $\times$ 7 items).}
    \label{fig:itemmap}
\end{figure}

The survey instrument used in the current study was developed based on the GenAI Literacy and Fluency framework \cite{dal2025scaffolding} (see Figure \ref{fig:matrix}). \textcite{dal2025scaffolding} adopted instrument design procedures by Wolfe and Smith \cite{wolfe2007instrument}, grounded in Messick's unified theory of validity \cite{messick1995validity}. Item construction was based on a process of defining a criterion-referenced framework \cite{griffin2007comfort}. Specifically, the item development blueprint used a combination of stems and response options written as observable behaviours in the framework. Items (n=7) in the GenAI Literacy and Fluency framework represent a single indicative behaviour for the key construct under investigation, that of AI literacy. Each of the 7 items has response options based on a hierarchical scale for an indicative behaviour that captures a level of proficiency from emerging to sophisticated guided by Bloom's revised taxonomy \cite{krathwohl2002revision} and coded from 1 to 4, along with an option to say none of the options apply (coded as 0). An example of an item and its corresponding response options for the \textit{Remember} cognitive dimension, covering all knowledge dimensions for this cognitive dimension in the framework, is presented in Figure \ref{fig:itemmap}.

This design aligns with \textcite{vygotskij1993fundamentals} who argued that transformations in performance quality should have what respondents do (performance indicators) and also how well each indicative behaviour is performed (criterion levels). Thus, an item (n=7) is defined as a single indicative behaviour (indicator) plus its corresponding criterion levels (criteria) (n=28). This method replaces common response option styles such as Likert rating scales, and instead describes response options as observable, self-assessable behaviours using accessible language. 

Each survey included 28 criteria thresholds across the 7 primary items, covering all cognitive and knowledge dimensions of the framework. Respondents were prompted to select their own level of competence within the hierarchy of response options at the time of undertaking the survey. This method was the most appropriate approach for the aims of a broader study within which the current study sits, that of targeting learning pathways for different learners. 

\subsection{Procedure}

The study was conducted between February and August (2025) at The University of Melbourne, Australia. Participants were recruited using a convenience sampling strategy. More specifically, advertisements were posted in the university's Learning Management System (Canvas), direct emails with requests to circulate the survey information were sent to academic departments, and announcements were posted in a university-wide `AI in Education' community of practice. 

This research was conducted under ethics approval from the University's Human Research Ethic Committee [Ethics ID \#31178]. All participants were provided with a Plain Language Statement detailing the study's purpose and procedures and gave their informed consent before beginning the survey. All survey data was anonymised prior to analysis to ensure participant confidentiality.

\subsection{Analysis}

The sample used for this study consisted of $n$ = 158 participants. The psychometric framework draws on two complementary methodologies. \textit{Rasch measurement theory}~\cite{masters1982rasch} provides a probabilistic model that places both item difficulty and person ability on a shared interval scale. For polytomous items with ordered response categories, the Partial Credit Model~\cite{masters1982rasch} estimates a separate difficulty parameter for each transition between adjacent response levels, here corresponding to the 28 criteria thresholds (4 levels $\times$ 7 items). Critically, the model generates \textit{fit statistics} indicating whether items and persons conform to the expected cumulative hierarchy; departures from fit signal response patterns inconsistent with a single linear continuum. Analysis was conducted in R using the \texttt{TAM} package~\cite{robitzsch2020package}. For Rasch calibration, the final sample size ($n=158$) exceeds the standard psychometric heuristic of $n \geq 100$ required to achieve stable item difficulty estimates within $\pm 0.5$ logits at 95\% confidence \cite{linacre1994sample}.

To investigate the stability of skill trajectories across groups, we prioritised \textit{Guttman ordering analysis}~\cite{griffin2014modified,guttman1950problem} over Differential Item Functioning (DIF), which requires large sample sizes to achieve adequate power. Guttman analysis offers a deterministic visual model robust to smaller cohorts, allowing granular inspection of individual response vectors. In a perfect Guttman scale, items arranged by difficulty and respondents arranged by total score produce a characteristic triangular pattern: mastery responses (coded~`1') cluster in the top-left, non-mastery (coded~`0') in the bottom-right. Any `1' appearing to the right of a `0' for the same respondent indicates a skill acquired ``out of sequence''---a deviation from the theoretical hierarchy revealing a non-linear learning trajectory. This makes Guttman charts well-suited for identifying ``jagged profiles'' at the subgroup level. Because Rasch models utilise probabilistic estimation to accommodate missing data, the full sample ($n$ = 158) was retained to establish stable item difficulty calibrations. Guttman scaling is a deterministic model requiring complete response vectors; all response vectors across the three subgroups were complete and fully retained for the Guttman analysis (students $n$ = 98, academic staff $n$ = 34, professional staff $n$ = 26).

In sequence, the two models address different questions: Rasch fit operates at the population level (RQ1), while Guttman disaggregates by subgroup to reveal cohort-specific trajectories (RQ2). To quantify agreement between cohorts, Spearman's rank correlations were calculated on each subgroup's item-difficulty ordering. Rather than null-hypothesis significance testing---conceptually misaligned for comparing finite ranked structures---we report 95\% confidence intervals computed via Fisher $z$-transformation to model uncertainty arising from the exploratory subgroup sample sizes.


\section{Results}

\subsection{Construct Dimensionality and Item Fit}
\label{itemfit}
To address RQ1, regarding whether GenAI literacy behaves as a uniform, unidimensional trait, the data was analysed using a Rasch model for partial credit scoring \cite{masters1982rasch}. First, at the item level, the analysis demonstrated strong model fit (see Table~\ref{tab:item_fit}). The weighted Mean Square (MNSQ) values for all items fell within the recommended bounds of 0.75 to 1.33  \cite{wright1994reasonable}. This indicates that the 28 criteria function consistently relative to one another---that is, the items collectively measure a shared domain of GenAI competency without individual items behaving erratically or contradicting the others. Therefore, the IRT Partial Credit Model showed that the items functioned coherently to indicate a single underlying construct and the strong item fit indicates that responses can be located along a common latent dimension. The data had different subgroups and allowed RQ2 to be investigated, thus reviewing relative ordering of item difficulty across groups. Differences in item hierarchy reflects variation in sequencing, or perceived difficulty among subgroups. The item-level evidence therefore supports the theoretical validity of the framework's content structure, while the person-level analysis addresses whether this shared domain is expressed uniformly across the population.

\begin{table}[ht]
    \centering
    \caption{Item fit to the Rasch model for the seven primary indicative behaviours}
    \label{tab:item_fit}
    \begin{tabular}{lc}
        \toprule
        \textbf{Item ID} & \textbf{Weighted MNSQ} \\
        \midrule
        Q1 & 1.07 \\
        Q2 & 1.12 \\
        Q3 & 1.09 \\
        Q4 & 0.94 \\
        Q5 & 0.92 \\
        Q6 & 0.94 \\
        Q7 & 0.92 \\
        \bottomrule
    \end{tabular}
\end{table}

However, although the \textit{items} function cohesively, analysis of \textit{person-level} fit revealed differences among groups in terms of predicting performance on a particular item. Person-level outfit MNSQ values ranged widely from 0.15 to 2.46, with a substantial proportion of respondents exhibiting outfit MNSQ $> 1.5$, indicating idiosyncratic, ``jagged'' response patterns requiring further investigation. 

Therefore, in response to RQ1, the data show that GenAI literacy does not behave as a single uniform latent trait throughout the diverse university ecosystem. The variance in person-fit and low internal consistency confirms that a ``one-size-fits-all'' psychometric ladder fails to capture the reality of learner behaviour. Because the assumption of a universal developmental pathway is violated at the population level, we must disaggregate the data. Acknowledging that the small sample size is a limitation, to undertake an exploration to address RQ2, we utilise subgroup-level Guttman analysis to isolate the specific, deterministic trajectories occurring within distinct academic cohorts.

\subsection{Order of Emergence and Skill Trajectories}

To investigate the divergent learning pathways signalled by the Rasch analysis (RQ2), we first constructed a cross-cohort endorsement heatmap, then examined the individual Guttman charts for each cohort to gain more detailed insights.

\begin{figure}[!b]
    \centering
    \includegraphics[width=\linewidth]{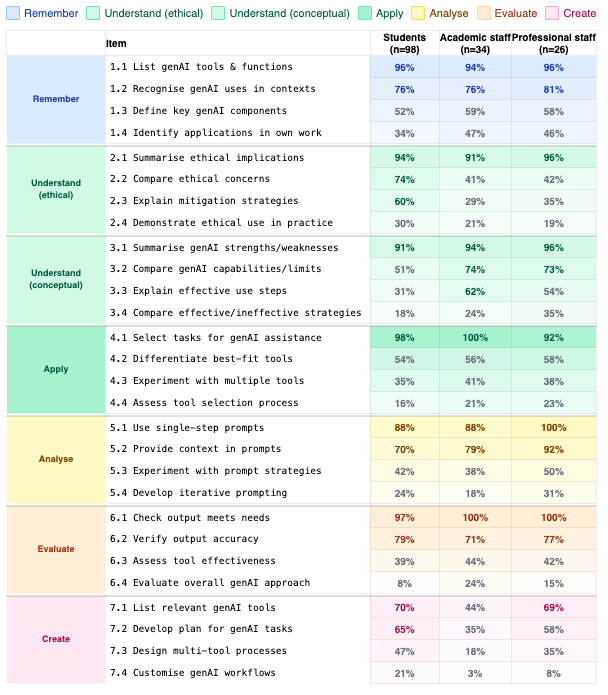}
    \caption{Cross-cohort endorsement heatmap for students, academics and professional staff. Items are grouped by cognitive dimension and colour-coded by GenAI Literacy and Fluency framework level. Each cell reports the percentage of respondents within a cohort who self-reported mastery of that item; cell saturation scales with endorsement rate such that darker cells indicate more widely shared competence.}
    \label{fig:heatmap}
\end{figure}

\begin{figure}[ht]
    \centering
    \includegraphics[width=0.8\textwidth]{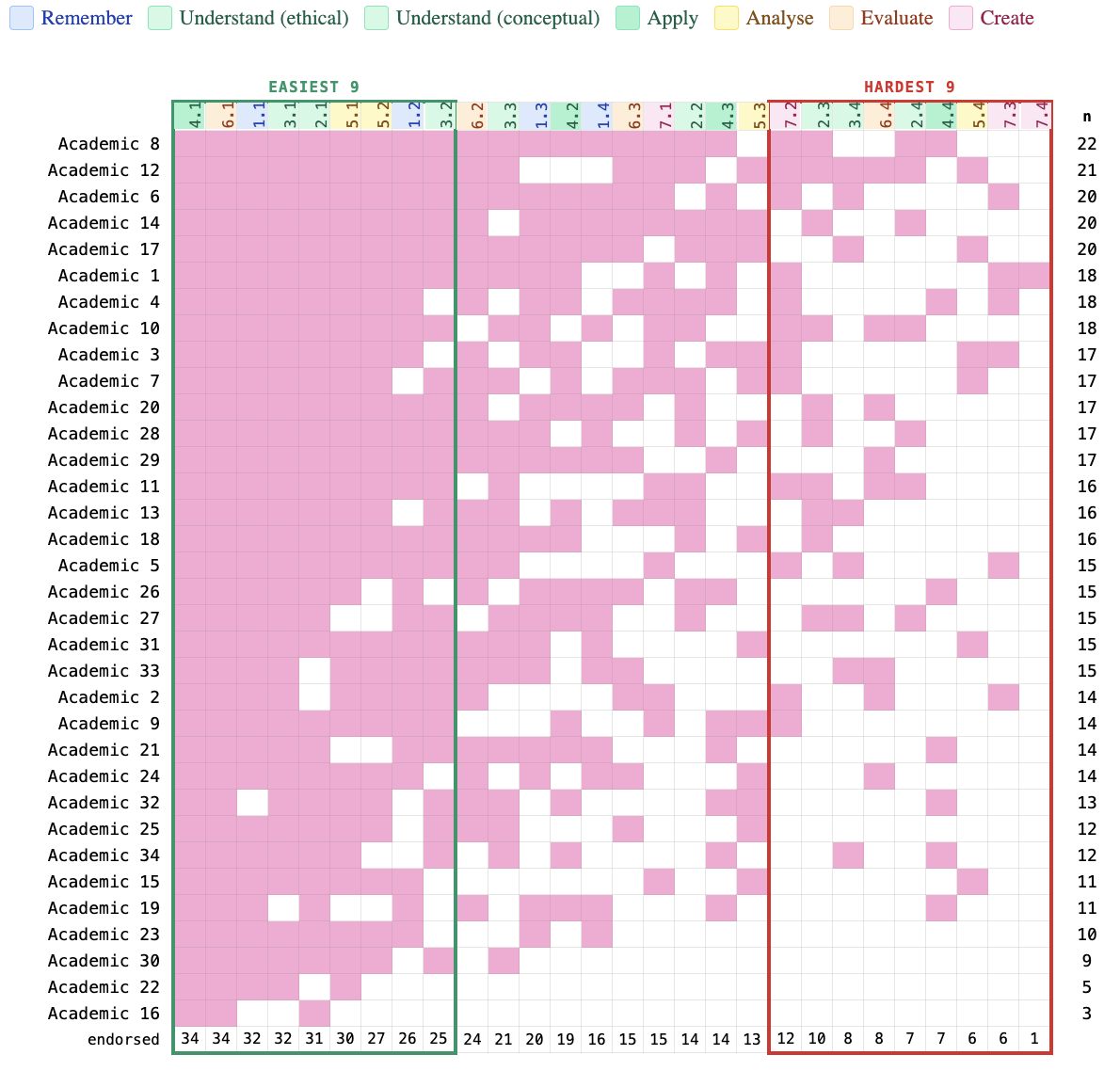} 
    \caption{Guttman chart for the academics ($n = 34$). Each item is coloured by its cognitive dimension group. Rows are individual respondents ordered by total score (highest to lowest). Pink cells indicate task endorsement (`1': can do); uncoloured cells indicate non-endorsement (`0': cannot yet do). The bottom row and rightmost column show item endorsement sums and individual totals, respectively. Lastly, an item-level polarity analysis was conducted, examining the easiest (most frequently endorsed) and hardest (least frequently endorsed) criteria for each cohort.}
    \label{fig:guttman_academic}
\end{figure}

\textbf{The Academic Staff}  display a trajectory broadly consistent with Bloom's taxonomy (see Figure~\ref{fig:heatmap} and Figure~\ref{fig:guttman_academic}). The most widely endorsed items are pragmatic entry points already embedded in scholarly workflows------selecting tasks for genAI assistance (item 4.1) and checking that outputs meet needs (item 6.1), both at 100\%---with the basic foundational items also highly endorsed (1.1, 2.1, and 3.1 all at 91--94\%). Create-level items occupy the lowest tier, bottoming out at item 7.4 (``Customise genAI workflows'') at just 3\%. This foundational-to-advanced gradient suggests that scaffolding for academics can build incrementally on existing evaluative competencies toward higher-order creation and workflow design.

\begin{figure}[ht]
    \centering
    \includegraphics[width=0.8\textwidth]{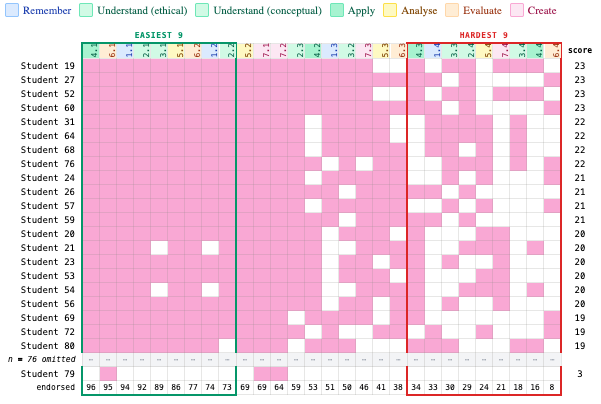}
    \caption{Guttman chart for the students ($n = 98$); plotting conventions as in Figure 4.}
    \label{fig:guttman_student}
\end{figure}

\textbf{The Student Cohort} reveals a striking inversion of the academic pattern (see Figure~\ref{fig:heatmap} and Figure~\ref{fig:guttman_student}). For visual clarity, Figure \ref{fig:guttman_student} shows the 21 highest-scoring students and the lowest-scoring student, with 76 intermediate rows omitted; footer counts reflect the full $n = 98$ sample. Create-level items 7.1 and 7.2 reach 70\% and 65\% endorsement---above several foundational items that would normally precede them. Item 1.3 (``Define key genAI components'') is endorsed by only 52\%, falling below both, while the most advanced threshold of each dimension sits lowest of all---bottoming out at item 6.4 (``Evaluate overall genAI approach'') at just 8\%, suggesting that metacognitive self-evaluation is largely absent even among active creators. Conceptual knowledge items present the lowest endorsement rates outside the hardest \textit{Evaluate} threshold: items 2.3 and 2.4 reach only 31\% and 18\% respectively, despite representing knowledge that would normally precede applied or creative use. Furthermore, 32 of the 33 students in the Guttman visualisation (97\%) showed at least one out-of-sequence mastery response, confirming that jagged, non-linear profiles are the norm here. This ``inverted Bloom's'' pattern---creation before conceptual consolidation---means interventions cannot assume a foundational base; they must work backwards from students' existing creative practice to build in the conceptual, ethical, and evaluative competencies it depends on.

\begin{figure}[ht]
    \centering
    \includegraphics[width=0.8\textwidth]{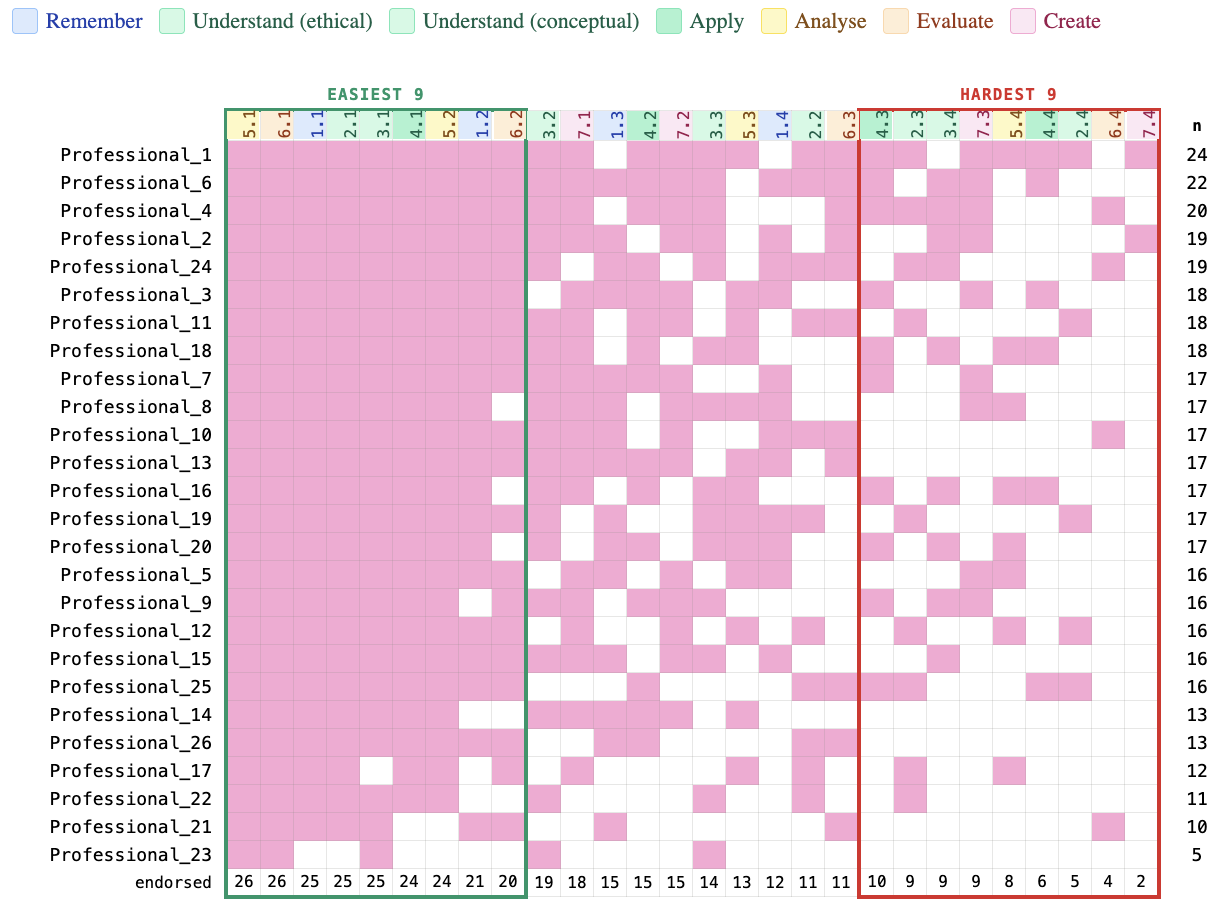}
    \caption{Guttman chart for the professional staff ($n = 26$); plotting conventions as in Figure 4.}
    \label{fig:guttman_professional}
\end{figure}

\textbf{The Professional Staff} occupy a middle ground and show distinct pragmatic priorities (see Figure~\ref{fig:heatmap} and Figure~\ref{fig:guttman_professional}). Operational items---using single-step prompts (5.1) and checking outputs (6.1)---are jointly most endorsed at 100\%, and basic ethical awareness is high (item 2.1, ``Summarise ethical implications,'' at 96\%), consistent with compliance-driven awareness under institutional policy. Create-level items span the widest range of any cohort, from basic tool planning (item 7.1, 69\%) down to advanced workflow customisation (item 7.4, 8\%). The resulting profile is pragmatically compressed rather than hierarchically ordered, so scaffolding here requires a different logic again: leveraging existing ethical and evaluative awareness while extending into the under-developed analytical and creative competencies.

\subsection{Comparative Analysis of Trajectories}
To address RQ2, regarding the stability of the order of emergence further, we first examine the specific item-level behaviours driving the divergence before statistically quantifying the overall trajectories.

To ground the visual divergences in observable behaviour, we inspected the easiest (most frequently endorsed) and hardest (least frequently endorsed) nine thresholds for each cohort, marked as the green and red bands in the Guttman charts (Figure~\ref{fig:guttman_academic}, Figure~\ref{fig:guttman_student}, Figure~\ref{fig:guttman_professional}). These bands expose the precise behavioural mechanics of the ``skill bypass.'' No \textit{Create}-level threshold appears in the easiest-9 of any cohort; conversely, \textit{Create} thresholds dominate the hardest-9 band for academics (three of four) and professional staff (two of four), but only one of four for students --- quantifying the developmental inversion across groups.

At the foundational level, all three cohorts share a pragmatic, transactional entry point into GenAI literacy. The easiest items universally endorsed across groups include selecting tasks appropriate for genAI assistance (Item 4.1) and evaluating whether outputs meet immediate needs (Item 6.1) --- the two most-endorsed items for both students and academic staff. Items 1.1, 1.2, and 5.1 are likewise consistently easy, occupying identical or near-identical ranks across all three cohorts. However, the developmental divergence becomes starkly apparent at the upper bounds of difficulty. For the \textit{Academic} cohort, the absolute most difficult thresholds are concentrated in advanced creation: designing structured workflows incorporating multiple tools (Item 7.3) and customising tools for specific goals (Item 7.4). This confirms their adherence to a traditional Bloom's trajectory, where complex creation is the final, most difficult frontier. This structured, foundational-first approach aligns with recent evidence that university educators adopt GenAI through a lens of systematic, perceived pedagogical usefulness rather than unstructured experimentation \cite{miranda2026exploring}. The \textit{Student} cohort hits a ``metacognitive ceiling.'' For students, high-level creation (Item 7.4, 21\%) is endorsed more frequently than the three hardest thresholds in their ordering: assessing their own decision-making process for tool selection (Item 4.4, 16\%), comparing effective/ineffective strategies (Item 3.4, 18\%), and evaluating their strategic approach (Item 6.4, 8\%).

\begin{figure}
    \centering
    \includegraphics[width=1\linewidth]{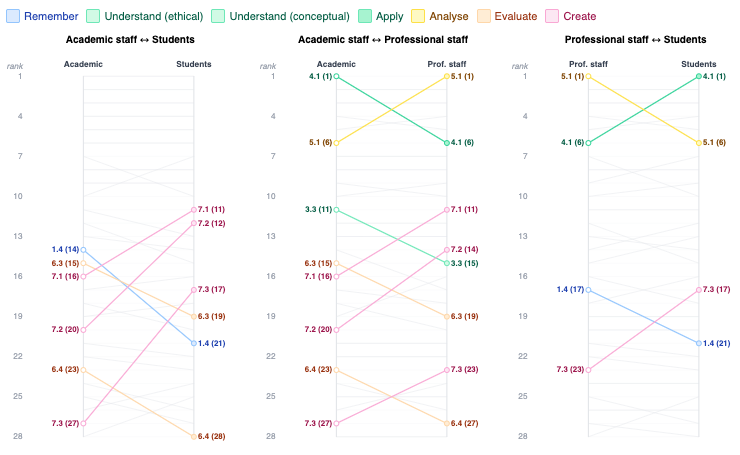}
    \caption{Pairwise rank-shift plots comparing item difficulty ordering across the three cohorts. Coloured lines (by cognitive dimension) indicate items shifting four or more rank positions between cohorts; grey lines indicate stable rankings.}
    \label{fig:ranking}
\end{figure}

Pairwise rank-shift plots comparing item difficulty ordering between cohorts reveal both similarities and differences (see Figure~\ref{fig:ranking}). \textit{Remember}-level and basic prompting items (1.1, 1.2, 5.1) are consistently easy across all groups. \textit{Create}-level items (7.1--7.3) rank substantially easier for students than for academic staff, who place them among the hardest items; professional staff occupy an intermediate position, finding these items more accessible than academics but less so than students. Conversely, contextualisation (1.4), ethical practice (2.3), and metacognitive evaluation (6.4) rank harder for students than for either staff cohort. Academic and professional staff diverge primarily in the creation domain, with professional staff finding basic creation planning (Item 7.1) as accessible as students, while academics perceive it as substantially more difficult.

To quantify the stability of these orders of emergence, we calculated Spearman's rank correlations on each subgroup's item-difficulty calibrations from the Guttman analysis ($N = 28$ thresholds; Table~\ref{tab:correlations}). Because the academic ($n = 34$) and professional staff ($n = 26$) subgroups are small, the resulting CIs do not propagate the calibrations' underlying measurement error, so these relational dynamics should be read as exploratory.

First, the moderate positive correlation between Academics and Professional Staff ($r = 0.524$, 95\% CI [0.187, 0.750]) suggests a shared institutional logic. The interval is entirely above zero, consistent with the interpretation that both staff cohorts appear to perceive skill difficulty through a similar lens, likely influenced by professional requirements for compliance, accuracy, and foundational understanding. While their absolute skill levels differ, their relative ranking of what constitutes a ``basic'' versus ``advanced'' task remains somewhat consistent. 

Second, the correlation between Professional Staff and Students ($r = 0.348$, 95\% CI [-0.029, 0.638]) is weak-to-moderate, with a wide interval that spans zero but skews positive. This suggests a tentative degree of overlap in pragmatic orientation --- professional staff and students share some common ground in how they rank accessible GenAI tasks --- but no consistent directional relationship can be firmly established from these exploratory data.

Third, the correlation between Academics and Students remains the weakest of the three ($r = 0.188$, 95\% CI [-0.199, 0.524]). The wide interval spanning both negative and positive values indicates no stable, shared developmental hierarchy between the two groups. While the point estimate is no longer negative, it remains non-significant and the interval includes values near zero in both directions, confirming that the order in which students self-assess skill mastery bears no consistent relationship to the order observed for academics. Educational interventions designed based on the academic's logic of progression (linear, concept-first) therefore cannot be assumed to resonate with the student's experience of progression (non-linear, output-first).

Finally, the thread that connects all cohorts is the lack of perceived metacognitive levels of knowledge in multiple cognitive domains. At the metacognitive level, individuals possess strategic knowledge and are able to self-reflect on their own understanding and identify gaps, a capability that is crucial for continuous and lifelong learning necessary with a dynamic and changing technology landscape.

\begin{table}[ht]
    \centering
    \caption{Item difficulty correlations by subgroup 
    ($N=28$ criteria thresholds)}
    \label{tab:correlations}
    \begin{tabular}{lccc}
        \toprule
        \textbf{Subgroup} & \textbf{Academic} & 
        \textbf{Professional} & \textbf{Student} \\
        \midrule
        Academic & 1 & 0.524 [0.187, 0.750] & 
        0.188 [-0.199, 0.524] \\
        Professional & & 1 & 0.348 [-0.029, 0.638] \\
        Student & & & 1 \\
        \bottomrule
        \multicolumn{4}{l}{\small{\textit{Note.} Brackets 
        indicate 95\% Confidence Intervals computed via 
        Fisher $z$-transformation.}}
    \end{tabular}
\end{table}

\section{Discussion}

This study was designed to address two research questions. RQ1 concerned whether the instrument's items function as indicators of a shared construct, and whether the population responds homogeneously. The items cohered to measure a single domain of GenAI competency, yet person-level analysis revealed highly idiosyncratic, ``jagged'' response patterns. This yields our first contribution: empirical evidence that GenAI literacy does not develop as a single, uniform trait across the university ecosystem, establishing the need to map distinct, role-based trajectories.

RQ2 concerned the developmental ``order of emergence'' of GenAI skills across cohorts, and what it reveals about their learning needs. Cohorts acquired skills in fundamentally different sequences: students self-assessed as proficient in high-level generative tasks while bypassing the foundational conceptual, evaluative, and ethical knowledge prioritised by academic and professional staff. This yields our second contribution: the formal identification of the ``skill bypass,'' and the empirical rationale for institutions to replace uniform curricula with targeted, diagnostic-driven interventions.

The mechanics of these divergent pathways are revealed through the psychometric tension between adequate item fit and high person-level misfit. Strong PCM item fit suggests the items measure the same overall construct. However, different groups may still experience or develop components of that construct in different orders. The hierarchy analysis therefore reflects differences in progression patterns, not necessarily different constructs. Limited data in each subgroup led to a novel approach to review item ordering. Our results, however, indicate a fundamental divergence in learning logic: while the \textit{Academic} cohort followed a more traditional and linear trajectory at the aggregate level (mastering ``Understanding'' before ``Creation''), the \textit{Student} cohort exhibited a structurally inverted profile of competence. Unlike traditional domains such as mathematics, where foundational skills almost invariably precede advanced application, our student data indicates that learners frequently perceived that they acquired ``advanced'' capabilities before consolidating several foundational conceptual, ethical, and evaluative competencies. For AI in education researchers, this signals that standard psychometric models which assume a single, linear continuum of ability fail to capture the reality of human-AI interaction. The implication is that future AI systems in education must move beyond linear scaffolding and instead account for the heterogeneity of learners and their skills by modeling learner capability as a dynamic, multi-dimensional construct that varies significantly by user role.

This structural inversion seen in the student data gives rise to the ``skill bypass,'' a phenomenon that carries significant pedagogical risks. By inverting Bloom's Taxonomy, the bypass allows users to self-report proficiency at Bloom's ``Create'' level without the prerequisite domain knowledge. This stands in sharp contrast to the \textit{Professional Staff} cohort, whose trajectory tended to prioritise regulatory compliance and ethical evaluation over creation at the aggregate level. While the student ``bypass'' offers pragmatic efficiency, it creates a dependency risk and a sense of competence without actual learning. In the cycle of fluency conceptualised by \textcite{dal2025scaffolding}, the cycle starts with the analysis phase, which leverages an individual's knowledge and understanding of GenAI concepts and ethical considerations. Although this functional fluency allows for rapid task completion, the underlying analysis often relies on flawed decision making, without a solid foundation of GenAI literacy. This flawed decision making could lead to poor problem solution fit, in other words, an inefficient or inappropriate allocation of tasks to GenAI. Without a solid underlying logic grounding the decision making process, students leveraging the skills bypass may become over-reliant and develop dependencies. This profile creates a dangerous isomorphism with the AI itself. As \textcite{galla2025generative} note, GenAI models also exhibit this ``inverted'' capability---excelling at surface-level synthesis while lacking the embodied, causal reasoning required for deep understanding. This risk is compounded by the tendency of current GenAI interfaces to passively mirror user intent. As \textcite{rinja2026unpacking} recently observed, when students adopt an executive, delegation-oriented stance, AI systems default to an ``executor'' role, mimicking and reinforcing passive behaviour rather than providing the pedagogical friction required for learning. This dynamic perfectly encapsulates the threat of ``metacognitive laziness'' \cite{fan2025beware}, where the mental effort required for self-regulation is offloaded to the machine. When an ``inverted'' student relies on an overly compliant, ``inverted'' AI, the human-AI loop becomes completely decoupled from verification. This illusion is dangerous because it collapses under scrutiny; without foundational understanding, the learner cannot debug, verify, or correct the AI teammate. Consequently, instruction must not stop students from ``creating'' early, but must recognise this capability as fragile, requiring educators to retrofit ``understanding'' into the workflow of a user who already feels fluent.

At an institutional level, addressing this divergence between cohorts demands a fundamental shift in how we design adaptive learning environments. The weak correlation ($r = 0.188$) between Students and Academics indicates that these groups operate in distinct developmental spaces. For higher education institutions, this disconnect presents a systemic risk: curricula designed by academics based on their own linear experience---or aligned with prescriptive, foundational-first models \cite{hutson2025scaffolded}---will likely fail to engage students who are already operating in an inverted paradigm. Similarly, the specific compliance needs of Professional Staff are not well-served by generalist ``AI for Education'' approaches. This problematises the ``one-size-fits-all'' delivery model for the entire institution. Instead, the distinct orders of emergence validated here argue for modular, diagnostic-driven interventions that respect the pragmatic context of each group. The instrument used in this study acts as a diagnostic dashboard, allowing learners (whether students, staff, or academics) to visualise their own unique profile and test out of basics they already grasp. This approach aligns with the heutagogical principle of self-determined learning \cite{blaschke2012heutagogy}, encouraging users to engage with content that matches their actual Zone of Proximal Development \cite{chaiklin2003zone}.

\subsection{Limitations}
This study has limitations. First, while the instrument measures self-perceived ability, perception does not always equate to performance. However, within a self-directed learning framework, perception remains a critical variable: if a user believes they are competent, they will not engage with foundational learning modules regardless of their actual skill gap. Second, the analytical sample ($n=158$) was insufficient for DIF analysis, though Guttman ordering provided a robust alternative for detecting deterministic patterns in smaller cohorts. Third, the data was collected from a single institution; future multi-institutional studies are required to confirm whether these distinct learning trajectories replicate across different geographic and cultural academic contexts. Future research should pair this diagnostic self-assessment with performance tasks to explore the correlation between the feeling of fluency and actual technical competence.
\section{Conclusion}
This study challenges the prevailing assumption that GenAI literacy develops along a single, linear trajectory. By subjecting a theoretical framework to empirical psychometric investigation, we revealed that the ``order of emergence'' for AI skills is highly sensitive to the user's role. While academics follow a more traditional pedagogical path, students frequently invert Bloom's taxonomy, prioritising high-level creation over foundational understanding. We term this phenomenon the ``skill bypass,'' which creates an illusion of fluency that educational institutions must address.

Although drawn from a single institution, these findings raise critical questions with international relevance for higher education. If this structural disconnect replicates across global contexts, it indicates a systemic challenge for the sector. For educators, students, administrators, and the AI in education community at large, these findings suggest that the path from ``tools to teammates'' is not uniform. A learner who creates without understanding is not a teammate, but a passenger. To foster genuine human-AI synergy, we must move beyond ``one-size-fits-all'' approaches and embrace diagnostic-driven architectures that respect the diverse, jagged profiles of our learners. 

The practical utility of this approach extends beyond theoretical redesign. Informed by the distinct developmental trajectories identified in this study, the host institution has already operationalised the 28-item framework into a suite of modular, diagnostic-driven GenAI training units for academic and professional staff. Initial deployment has demonstrated immense scalability and demand, with over 5,000 thousand enrolments and over 3,100 module completions recorded within the first nine months of release. Future research will systematically investigate these staff interaction patterns and satisfaction metrics, while expanding the modular architecture to the student cohort. Ultimately, bridging the gap between perceived fluency and actual performance is essential to ensuring that the agency we cultivate in learners is grounded in genuine, robust competence.

\section*{Acknowledgements}
This project was supported by the GEM Scott Teaching Fellowship (2024). We are grateful for the institutional support provided by the University of Melbourne and would like to acknowledge the collaborative contributions of colleagues across educational design, AI research, and ethics who provided expert feedback during the development and validation of the self-assessment framework.
\printbibliography[title={References}]

\end{document}